\documentclass[runningheads]{llncs}

\usepackage[utf8]{inputenc}
\usepackage[T1]{fontenc}
\usepackage{graphicx}
\usepackage{algorithm,algpseudocode}

\usepackage{url,hyperref} 
\usepackage{booktabs,array}
\usepackage{amsmath,amssymb}
\usepackage{comment}
\usepackage{color}

\begin{document}


\title{Large-scale multi-objective influence maximisation with network downscaling}

\titlerunning{Large-scale multi-objective influence maximisation}

\author{Elia Cunegatti\inst{1,2}\orcidID{0000-0002-1048-0373} \and
Giovanni Iacca\inst{1}\orcidID{0000-0001-9723-1830} \and
Doina Bucur\inst{2}\orcidID{0000-0002-4830-7162}}

\authorrunning{E. Cunegatti et al.}

\institute{
	University of Trento, Italy\\
	\email{elia.cunegatti@studenti.unitn.it, giovanni.iacca@unitn.it}
	\and
	University of Twente, The Netherlands\\
	\email{d.bucur@utwente.nl}
}
\maketitle 
\begin{abstract}
Finding the most influential nodes in a network is a computationally hard problem with several possible applications in various kinds of network-based problems. While several methods have been proposed for tackling the influence maximisation (IM) problem, their runtime typically scales poorly when the network size increases. Here, we propose an original method, based on network downscaling, that allows a multi-objective evolutionary algorithm (MOEA) to solve the IM problem on a reduced scale network, while preserving the relevant properties of the original network. The downscaled solution is then upscaled to the original network, using a mechanism based on centrality metrics such as PageRank. Our results on eight large networks (including two with $\sim$50k nodes) demonstrate the effectiveness of the proposed method with a more than 10-fold runtime gain compared to the time needed on the original network, and an up to $82\%$ time reduction compared to CELF.
\end{abstract}


\section{Introduction}
\label{sec:intro}
Given a social network for which the network structure is known and the process of influence propagation can be modelled, the problem of influence maximisation (IM) \cite{Richardson2003TrustMF} in its simplest form aims to select a certain number of participants (nodes) in the network, such that their combined influence upon the network is maximal. This is a combinatorial optimisation task, NP-hard for most propagation models \cite{kempe2003maximizing}. Various metaheuristics have been proposed to solve this problem, including (but not limited to) simulated annealing \cite{Jiang2011SimulatedAB}, 
genetic algorithms \cite{Bucur2016InfluenceMI,lotf2022improved}, memetic algorithms \cite{Gong2016AnEM}, particle swarm optimisation \cite{Gong2016InfluenceMI}, and, more recently, evolutionary deep reinforcement learning \cite{ma2022influence} and multi-transformation evolutionary frameworks \cite{wang2022multi}.  Multi-objective formulations of the IM problems have also been tackled, for instance in \cite{Bucur2017MultiobjectiveEA,bucur2018improving}.
In all cases, the drawback of these methods is their long runtime, which makes them infeasible to use on large social networks with more than $10^5$ nodes.

Here, aim to address precisely this computational issue. As in \cite{Bucur2017MultiobjectiveEA,bucur2018improving}, we consider a multi-objective formulation of the IM problem, where the two competing objectives are the number of selected nodes (to be minimised), and the combined influence (to be maximised). With respect to the previous literature, we contribute a scalable method for the multi-objective IM problem, which allows to tackle the problem for social networks orders of magnitude larger than before. The method is built around a multi-objective evolutionary algorithm (MOEA), but additionally employs a first step of \emph{graph summarisation}~\cite{Liu2016GraphSM} which downscales the network (by a configurable factor) while preserving its important structural properties, and a last step which upscales the solutions from the downscaled network to the original one. This approach allows the MOEA to run in feasible time, since it is executed on a smaller instance of the problem.

The runtime has always been a key issue in the literature on the IM problem.
Some previous works have tried to overcome this limitation by improving directly the effectiveness of the algorithm, see \cite{Li2018InfluenceMO} for a survey on this topic.
Some recent works have tried to minimise the runtime in billion-scale networks \cite{Li2019TipTopE,Wu2018ATS,Gney2021LargescaleIM}. Yet, to the best of our knowledge no previous work has attempted to tackle the problem by focusing on the input (i.e., the network), instead of the algorithm.

We tested our method on six different networks, using two propagation models.
For the downscaling process, we used three different values of scaling factor, to analyse how this affects the performance of our method.
For the upscaling process, we evaluated different centrality metrics. Finally, we tested our method on two large networks with high scaling factor values and compared the results with a classical heuristic algorithm. The results show that our method can achieve near-optimal or even better results, compared to the MOEA on the corresponding unscaled networks, while drastically reduce the runtime required.

The rest of the paper is structured as follows. In the next section, we describe our method. In Section \ref{sec:results}, we present the numerical results. Finally, we give the conclusions in Section \ref{sec:discussion}.


\section{Method}
\label{sec:methods}

\begin{figure}[ht!]
\includegraphics[width=\linewidth]{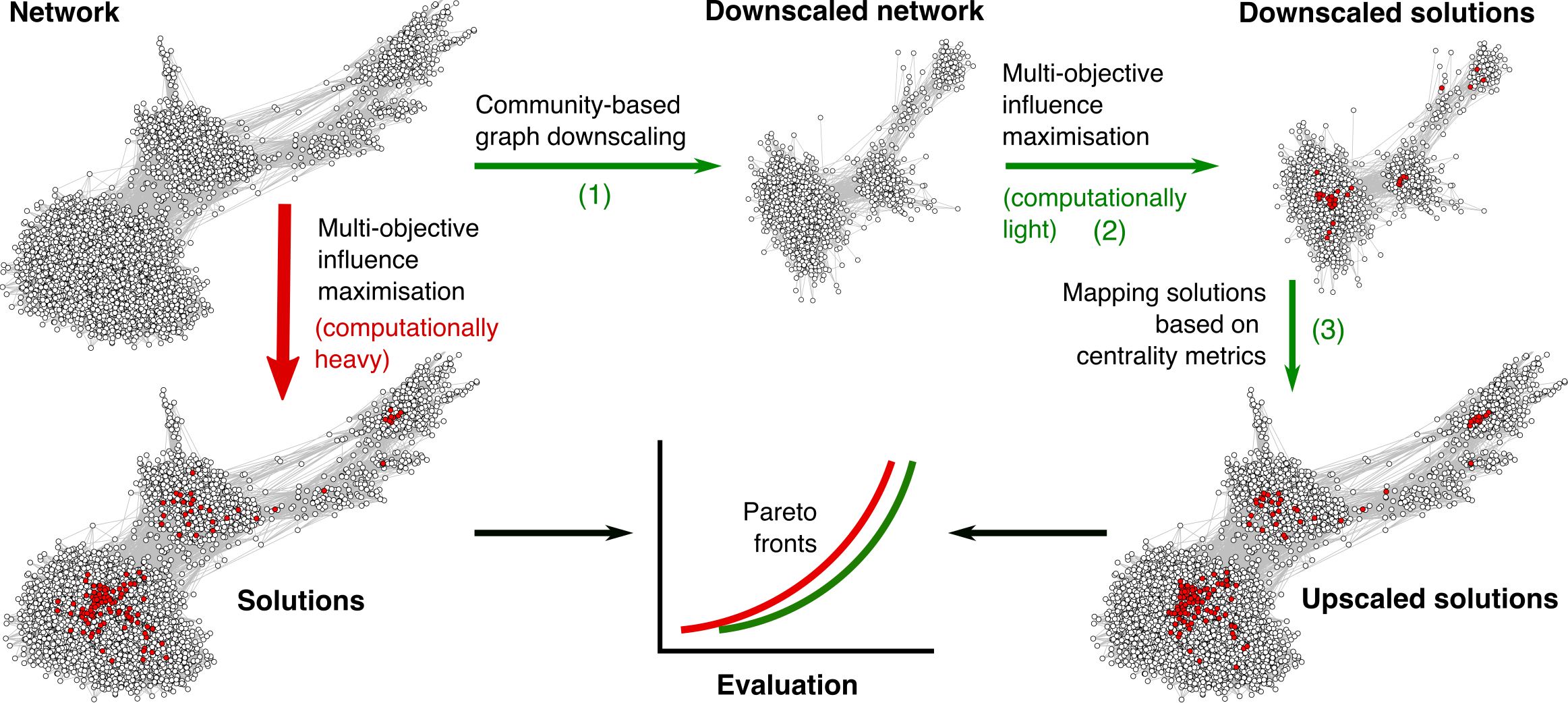}
\caption{The method. The larger the original network (top left), the less feasible IM is (bottom left). Our efficient three-step method scales down the network, does IM on the scaled network, then upscales the solutions (right).}
\label{fig:method}
\vspace{-0.2cm}
\end{figure}

We provide a first overview of the method in Fig.~\ref{fig:method}. For the MOEA, we use Non-Dominated Sorting Genetic Algorithm-II (NSGA-II) \cite{Deb2002AFA}, which has shown good results on this problem in prior work \cite{Bucur2017MultiobjectiveEA,bucur2018improving,iacca2021evolutionary}, but whose runtime made it prohibitive on large networks. This computationally heavy method is marked in Fig.~\ref{fig:method} on the left with a heavy red arrow. Instead of attempting to further improve the algorithm (with likely minor gains in efficiency), here we design an alternative which fundamentally changes the way we treat the problem input.

Given a large social network (top left in Fig.~\ref{fig:method}), in step (1) we synthesise a downscaled version of the network, for which the scaling factor is configurable. A network scaled by a factor $s=2$ would contain half of the nodes of the original, but otherwise preserve all the important properties of the original network: the number of communities is maintained constant, and the node degree distribution is scaled proportionally with the network size. In step (2), we apply the MOEA on this downscaled network, obtaining a number of non-dominated solutions (where each solution is a \emph{seed set}, i.e., the set of nodes from which influence is propagated). Only one such seed set is shown in Fig.~\ref{fig:method}, with the corresponding nodes marked in red. Finally, for step (3) we devise a method based on node centrality metrics to select the seed set in the original network, such that its nodes correspond (in terms of position in the network) to the ones found in step (2). Steps (1-3) output a Pareto front (PF) of solutions. To evaluate how accurate this method is, we also execute the MOEA on the original network, and compare the two PFs. In the rest of this section, we detail the three aforementioned steps\footnote{Code available at: \url{https://github.com/eliacunegatti/Influence-Maximization}.}.

\subsection{Step (1): Community-based downscaling}

The output of the downscaling process is a completely synthetic network, not identical to any part of the original network, but which appears very similar, yet much smaller. The \emph{scaling factor} $s$ is the key parameter when downscaling: we experiment with values in a geometric sequence, $s\in \{2, 4, 8\}$. The downscaling process satisfies the following requirements:
\begin{enumerate}
	\item preserves the \emph{number of communities} in the network;
	\item downscales the \emph{number of nodes and edges} by a factor $s$;
	\item preserves the \emph{node degree distribution}.
\end{enumerate}

We pose the last requirement because dynamic phenomena on a network (such as information propagation) have outcomes which depend on the node degrees. We use the Leiden algorithm \cite{Traag2019FromLT}, a state-of-the-art, scalable method for community detection. The algorithm is stochastic, so variations in the number and size of the communities are possible; because of this, we select one solution, then filter out from the network any communities that are too small (i.e., those that contain a number of nodes lower than the scaling factor $s$). We then preserve this number of communities obtained on the original network, downscaling (proportionally with $s$) the number of nodes and edges in each community. Per community, the shape of the node degree distribution is also preserved, as follows. Take the number of nodes in some original community to be $N$. We take a number of samples of size $N/s$ from the original node degrees of each community, then retain only the samples with the smallest Euclidean distance (computed in terms of mean degree and std. dev. per community) compared to the original ones. We thus have a desired node degree distribution for a downscaled network.

The downscaled network is then generated by the Stochastic Block Model (SBM), a random generative network model originally proposed in \cite{Holland1983StochasticBF}. This generates random networks with a configured number of communities, number of nodes per community, and edge density per community. Here, we used a more fine-grained, recent SBM method \cite{Karrer2011StochasticBA,Peixoto2017NonparametricBI} which takes also into account node degrees, and is implemented in the Python \texttt{graph-tool} library\footnote{\url{https://graph-tool.skewed.de}}.

\subsection{Step (2): MOEA on two objectives (cascade size and seed set size)}

Single-objective IM is the problem of finding those $k$ nodes with maximum collective influence upon the network. As mentioned above, a candidate solution for this is a \emph{seed set} (a set of node identifiers) of size $k > 0$. The multi-objective formulation in this study aims to maximise the collective influence, while minimising the size of the seed set $k$. In other words, the fitness of a candidate solution is a tuple of: (a) the {\bf estimated collective influence} of the seed set, and (b) the {\bf size of the seed set}. Both values are normalised with respect to the network size, to allow for a fair comparison between scaled and unscaled networks, as well as between networks of different sizes. We set the maximum possible value of $k$ to be 2.5\% of the network size.

As for the estimated collective influence, we model influence cascade using two classic, discrete-time propagation models for social networks \cite{kempe2003maximizing}. They simulate the dynamics of information adoption in a network modelled by graph $G$, in which a set $S$ of ``seed'' nodes are the initial sources of information. At any given time, the nodes in $G$ are in one of two states: ``activated'' (if they received the information and may propagate it further) or not. Initially, only the nodes in $S$ are activated. The information propagates via network links probabilistically: a probability $p$ models the likelihood of a source node activating a neighbouring destination node via their common link. The important quantity is the number of nodes eventually activated, also called the \emph{cascade size}---for IM, this is typically the main objective.

Algorithm~\ref{alg:cascade} gives a general view of cascade propagation models: set $A$ consists of all the activated nodes, and is initially equal to the seed set $S$. At each time step, recently activated source nodes try to activate their neighbours independently. If an activation fails, it is never retried (a destination node is assumed to have made its decision). If it succeeds, the propagation may continue; the process stops when no new nodes were activated in a time step. 

We use two model variants: {\bf Independent Cascade} (IC) and {\bf Weighted Cascade} (WC). IC was first introduced in marketing, to model the complex effects of word-of-mouth communication \cite{Goldenberg2001TalkOT}. In IC, the probability $p$ is equal across all links (when a node has more than one neighbour, their activations are tried in arbitrary order). WC further models the fact that a node's attention is limited: the probabilities of activation on links leading to a destination node $m$ are not uniform, but inversely proportional to the number of such links in $G$ (in other words, the degree of $m$), $p=1/deg(m)$ \cite{kempe2003maximizing}, where $deg(m)$ is the in-degree of node $m$, i.e., the number of edges incoming to $m$.

\begin{algorithm}[ht!]
	\caption{Cascade propagation models. $G$ is the graph, $S$ the seed set, and $p$ the probability that a link will be activated.} 
	\label{alg:cascade}
	{\small
		\begin{algorithmic}[1]
		 \Require $G$, $S$, $p$ 
		 \State $A \leftarrow S$ \algorithmiccomment{The complete set of activated nodes}
		 \State $B \leftarrow S$ \algorithmiccomment{Nodes activated in the previous iteration}
		 \While {$B$ not empty}
		 	\State $C \leftarrow \emptyset$ \algorithmiccomment{Nodes activated in the current iteration}
		 	\For {$n$\;in\;$B$} 
				\For{$m$\;in\;$neighbours(n) \setminus A$}
					\State $C \leftarrow m$ with probability $p$ \algorithmiccomment{Activation attempt}
				\EndFor
			\EndFor
			\State $B \leftarrow C$ and $A \leftarrow A \cup B$
		 \EndWhile
		 \State \Return $|A|$ \algorithmiccomment{The final size of the cascade}
		 \end{algorithmic} 
	 }
\end{algorithm}

Although a single execution of Algorithm~\ref{alg:cascade} is polynomial in the size of the network, the model is stochastic, and computing the expected cascade size exactly for a given seed set $S$ is \#P-complete \cite{Wang2012ScalableIM}. However, good estimates of $|A|$ can be obtained by Monte Carlo simulations: in our experiments, we run $100$ repetitions of Algorithm~\ref{alg:cascade} for each estimation. 

Concerning the MOEA, we used the implementation and parameterisation of NSGA-II adopted in prior work \cite{bucur2018improving,iacca2021evolutionary}. In short, the parent solutions are selected by tournament selection; the child solutions are generated by one-point crossover and random mutation. An archive keeps all the non-dominated solutions found, i.e., the PF. The replacement mechanism selects non-dominated solutions by their dominance levels, and then sorts them by crowding distance to prefer isolated solutions and obtain a better coverage of the PF.

To improve the convergence of the MOEA, we apply a \emph{smart initialisation} of its initial population, as proposed in \cite{Konotopska2021GraphawareEA}. First, we apply node filtering, which computes the influence of each node in the network separately, and then keeps the 50\% most influential nodes. Then, each of these nodes is added to a candidate solution with a probability proportional to its degree.
We summarise the parameters of the method in Table~\ref{tab:params}.

\begin{table}[ht!]
\vspace{-0.3cm}
\caption{Parameters of the method.}
\label{tab:params}
\vspace{-4mm}
\centering
	\begin{minipage}[t]{7cm}
	\begin{tabular}[t]{lr}
		\hline
		\textbf{Network parameters} & \\
		\hline
		Scaling factor $s$		& $\{2, 4, 8\}$ \\
		Max. seed set size $k$	& 2.5\% $\cdot$ network size \\
		IC probability $p$		& 0.05 \\
		No. simulations			& 100 \\
		\hline
	\end{tabular}
	\end{minipage}
	\begin{minipage}[t]{5cm}
	\begin{tabular}[t]{lr}
		\hline
		\textbf{NSGA-II parameters} & \\
		\hline
		Population size		& 100 \\
		Generations			& 1000 \\
		Elites				& 2 \\
		Crossover rate		& 1.0 \\
		Mutation rate		& 0.1\\
		Tournament size		& 5 \\
		\hline
	\end{tabular}
	\end{minipage}
\vspace{-11mm}
\end{table}

\subsection{Step (3): Upscaling}

Once the MOEA has been run on the downscaled network, the last step is to map the solutions back to the original network. This step takes in input two graphs (the original $G$ and the downscaled $G_s$) and a set seed on $G_s$, denoted as $S_s$. The task is to translate $S_s$ into an seed set $S$ on $G$. 

We achieve this by \emph{matching} nodes between the two graphs, based on their \emph{node centrality indicators}, namely node statistics which capture the position of the node in the network. We test the following classical centrality indicators, based on them being shown to be predictive of the node's influence \cite{bucur2020top}: each node's \emph{degree}, \emph{eigenvector centrality} and its variants \emph{PageRank} with a 0.85 damping factor and \emph{Katz centrality}, \emph{closeness}, \emph{betweenness}, and \emph{core number} (see \cite{newman2018networks} for their definitions). 

For a seed set $S_s$ in $G_s$, we find a matching seed set $S$ of $|S_s| \times s$ nodes in $G$. We do this per community. Each node in $S_s$ has a \emph{rank} in its community, based on the centrality values of all nodes in that community. We then search in $G$ (among the nodes in the corresponding community) for $s$ nodes with the most similar ranks. These nodes form $S$.

{\bf Evaluation.} We evaluate the PFs obtained, particularly to compare between the MOEA results on the original network and those obtained with our new method. We use the \textit{hypervolume} (HV) indicator (also known as Lebesgue measure, or S metric) proposed in \cite{zitzler1998multiobjective}. This is calculated as the volume (of the fitness space) dominated by each solution in the PF with respect to a reference point. The \textit{hyperarea} (HR) \cite{zitzler1998multiobjective} is the ratio of two HVs, and is used here in the final evaluation step (bottom center in Fig.~\ref{fig:method}). 


\section{Results}
\label{sec:results}

{\bf Network data.} We test our method on six real-world social network topologies (listed in Table~\ref{tab:dataset}). These range between 4 039 and 28 281 nodes, with variable average degrees (and thus network densities), and variable number and size of communities. {\bf Ego Fb.} denotes data merged from many ego networks on Facebook, collected from survey participants at a large university. {\bf Fb. Pol.} is a network of mutually liked, verified politicians' pages on Facebook. {\bf Fb. Pag.} is similar, but with Facebook pages from various categories. {\bf Fb. Org.} is a network of friendships among Facebook users who indicated employment at one corporation. {\bf PGP} is the largest connected component in the network of PGP encryption users. {\bf Deezer} represents online friendships between users of the Deezer music platform. All graphs are undirected and connected.

\begin{table}[ht!]
\vspace{-0.3cm}
\caption{Networks considered in the experimentation.}
\label{tab:dataset}
\centering
	\begin{tabular}{p{2cm} rr rrr | rrr}
	\hline
	&&&\multicolumn{3}{c}{\textbf{Communities}}&\multicolumn{3}{c}{\textbf{Node degrees}}\\
	\cline{4-9}
	\textbf{Network} & $\quad$\textbf{Nodes} & $\quad$\textbf{Edges} & $\quad$\textbf{Num.} & $\quad$\textbf{Min.} & $\quad$\textbf{Max.} & $\quad$\textbf{Avg.} & $\quad$\textbf{Std.} & $\quad$\textbf{Max.} \\ \hline
	Ego Fb. \cite{McAuley2012LearningTD} 			& 4 039 & 88 234 &17&19& 548& 43.90 & 52.41 &1045\\ 
	Fb. Pol. \cite{rozemberczki2019gemsec}			& 5 908 & 41 729 &31&8&562& 14.12 & 20.09 &323\\ 
	Fb. Org. \cite{Fire2013OrganizationMU}			& 5 524 & 94 219 &13&35&1045& 34.11 & 31.80 &417\\ 
	Fb. Pag. \cite{rozemberczki2019gemsec}			& 11 565 & 67 114 &31&8&1916& 11.60 & 21.28 &326\\ 
	PGP \cite{boguna2004models}						& 10 680 & 24 316 &91&8&668& 4.55&8.07 &205\\ 
	Deezer \cite{rozemberczki2020characteristic}	& 28 281 & 92 752 &71&8&4106& 6.55 &7.94 & 172\\ \hline
	\end{tabular}

\end{table}

In the remainder of this section, we experiment with and evaluate our method. We present results for the three distinct steps of the method (as per Fig.~\ref{fig:method}).

\subsection{Community-based downscaling of large networks}
\begin{figure}[ht!]
\centering
	\includegraphics[width=0.9\textwidth]{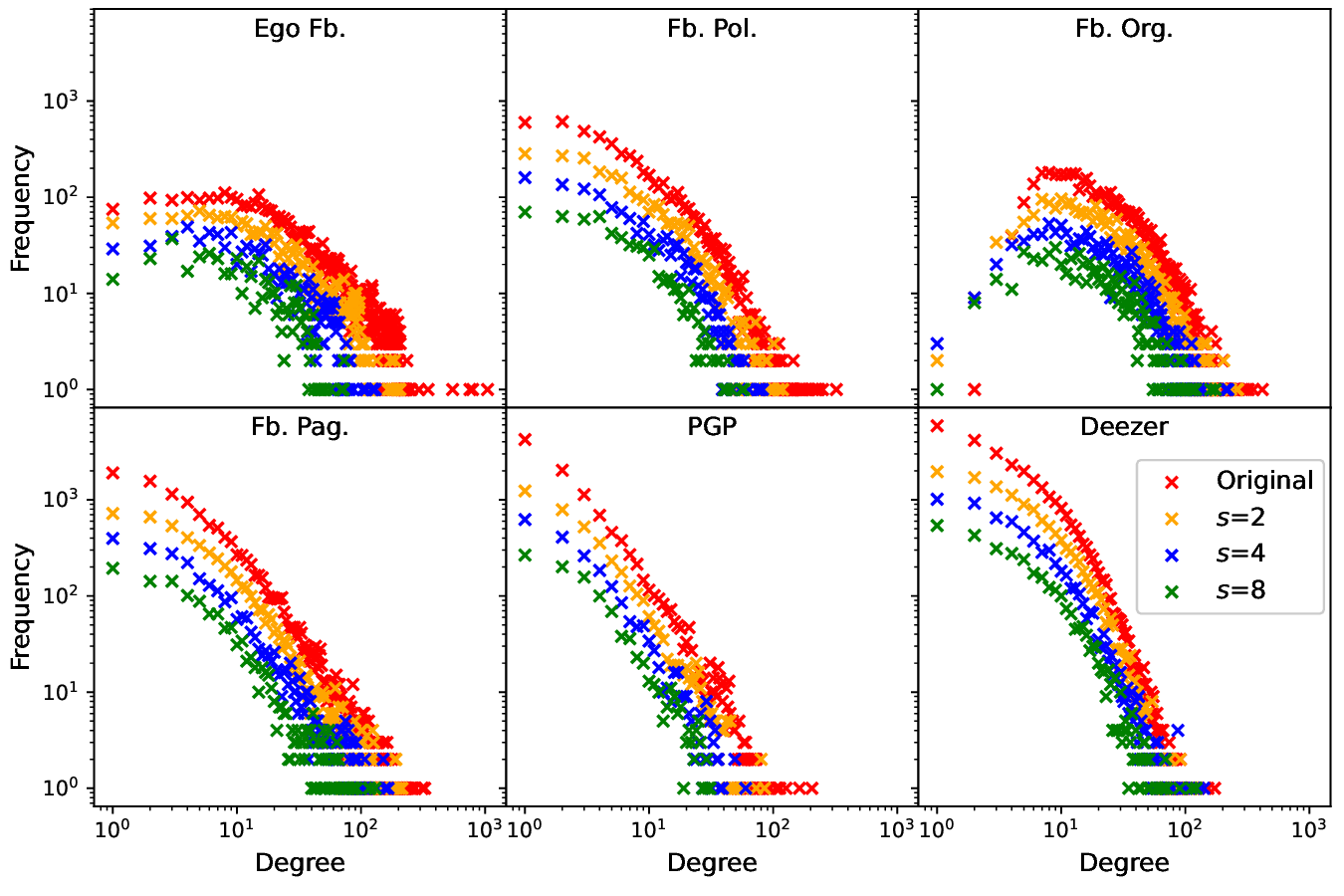}
	\caption{Degree distributions, before and after downscaling.}
\label{fig:scaling}
\end{figure}

This step obtains synthetic scaled networks, with the scaling factor $s$. These networks have the same number of communities, a scaled number of nodes and edges, and the same shape of the degree distribution. For $s \in \{2,4,8\}$, we show in Fig.~\ref{fig:scaling} the degree distributions (in log-log scale) for the six networks: that of the original (unscaled) network, and that of synthetic, scaled versions. Fig.~\ref{fig:method} included plots of Fb. Org. before and after downscaling ($s=4$). 

The networks plausibly fit typical power-law degree distributions, in which the fraction of nodes with a certain degree $d$ is proportional to $d^{-\alpha}$ where $\alpha$ is positive (so, decreases with $d$), after a cutoff point $d_\text{min}$. In real-world networks from various domains, the power-law parameter $\alpha$ is often measured between 1.5 and 3 \cite{broido2019scale}, which is mostly the case also for these six social networks. This parameter can be seen in Fig.~\ref{fig:scaling}, in the linear slope of all distributions (for high enough degrees). Our downscaling step preserves the power-law parameter $\alpha$ between the original and downscaled networks, naturally while scaling down the degree frequencies with $s$. We show the fitted values for $\alpha$ in Table~\ref{tab:powerlaw}, from which it is clear that the downscaling method introduced little error in the shape of the degree distribution.

\begin{table}[ht!]
\centering
\caption{Degree distribution preservation, power-law coefficients.}
\label{tab:powerlaw}
	\begin{tabular}{lrrrr}
	\hline
	& {\bf Original} & $\quad s=2$ & $\quad s=4$ & $\quad s=8$ \\
	\hline
	Ego Fb. & 1.32 & 1.35 & 1.38 & 1.42 \\
	Fb. Pol. & 1.50 & 1.52 & 1.54 & 1.57 \\
	Fb. Org. & 1.40 & 1.32 & 1.33 & 1.35 \\
	Fb. Pag. & 1.60 & 1.58 & 1.60 & 1.62 \\
	PGP & 2.11 & 1.94 & 1.97 & 1.93 \\
	Deezer & 1.73 & 1.66 & 1.68 & 1.68 \\
	\hline
	\end{tabular}
	\vspace{-5mm}
\end{table}

In the next step, we run the MOEA on the downscaled networks (and, for evaluation, also on the original networks).

\subsection{MOEA and solution upscaling: the optimality of solutions}

\begin{figure}[ht!]
\vspace{-0.3cm}
\centering
	\includegraphics[width=\textwidth]{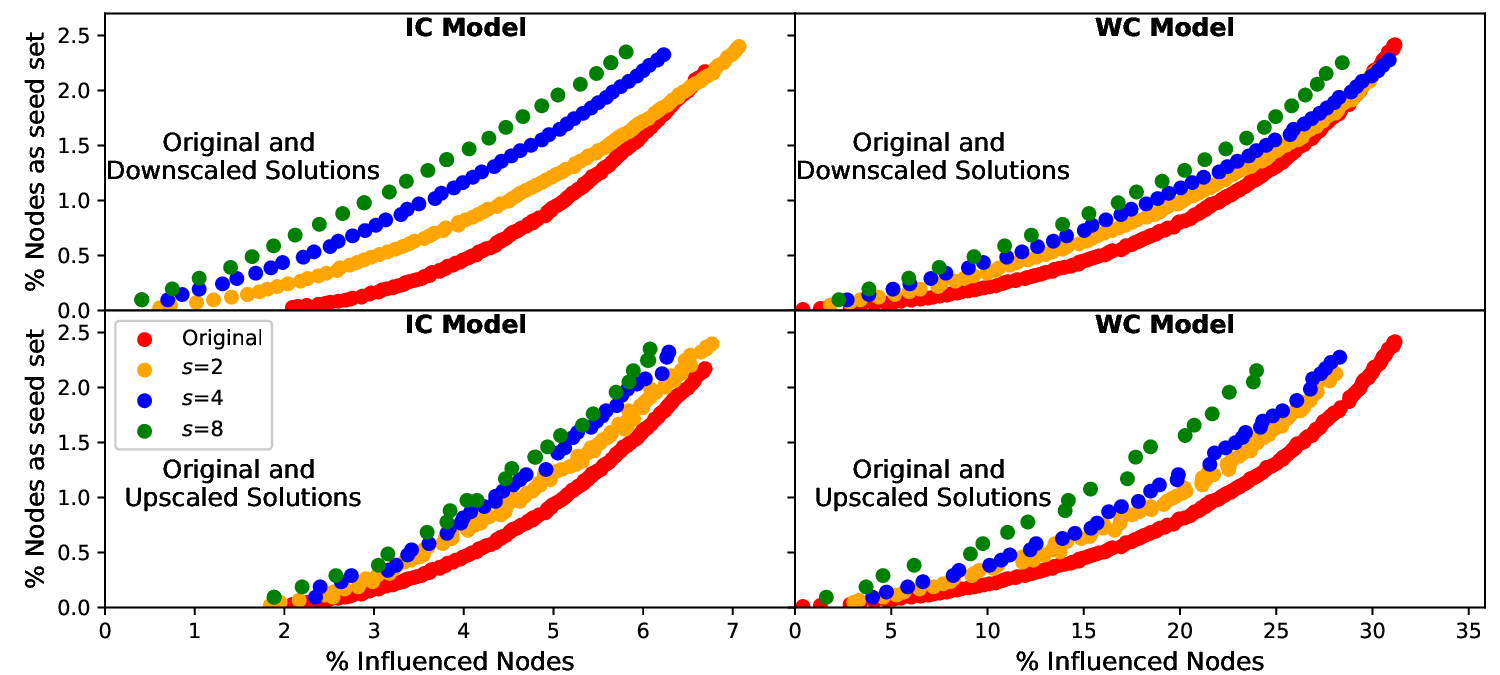}
	\caption{PGP Pareto fronts: (top) for the original and downscaled networks, and (bottom) for the original network and the upscaled solutions.}
\label{fig:PF1}
\end{figure}

The optimisation process obtains a two-dimensional PF of solutions, where as said each solution is a seed set of $k$ nodes of the network. We show an example run (randomly selected out of the 10 performed) of PFs in Fig.~\ref{fig:PF1}, for the PGP case and for both propagation models (IC and WC), obtained by the MOEA on the original network and on the downscaled networks (for three values of $s$). The PFs for the original and downscaled networks (top in Fig.~\ref{fig:PF1}) show that running the optimisation on the downscaled networks preserves the shape of the PF, but slightly lowers the values reached for the main objective, namely the percentage of influenced nodes (i.e., the cascade size). The more downscaled the network is, the more pronounced this effect appears to be. When upscaling the solutions obtained on the downscaled networks, here using the PageRank centrality in the upscaling process (bottom in Fig.~\ref{fig:PF1}), this gap closes for IC, but not for WC propagation.

\begin{figure}[ht!]
\vspace{-0.2cm}
\centering
	\includegraphics[width=\textwidth]{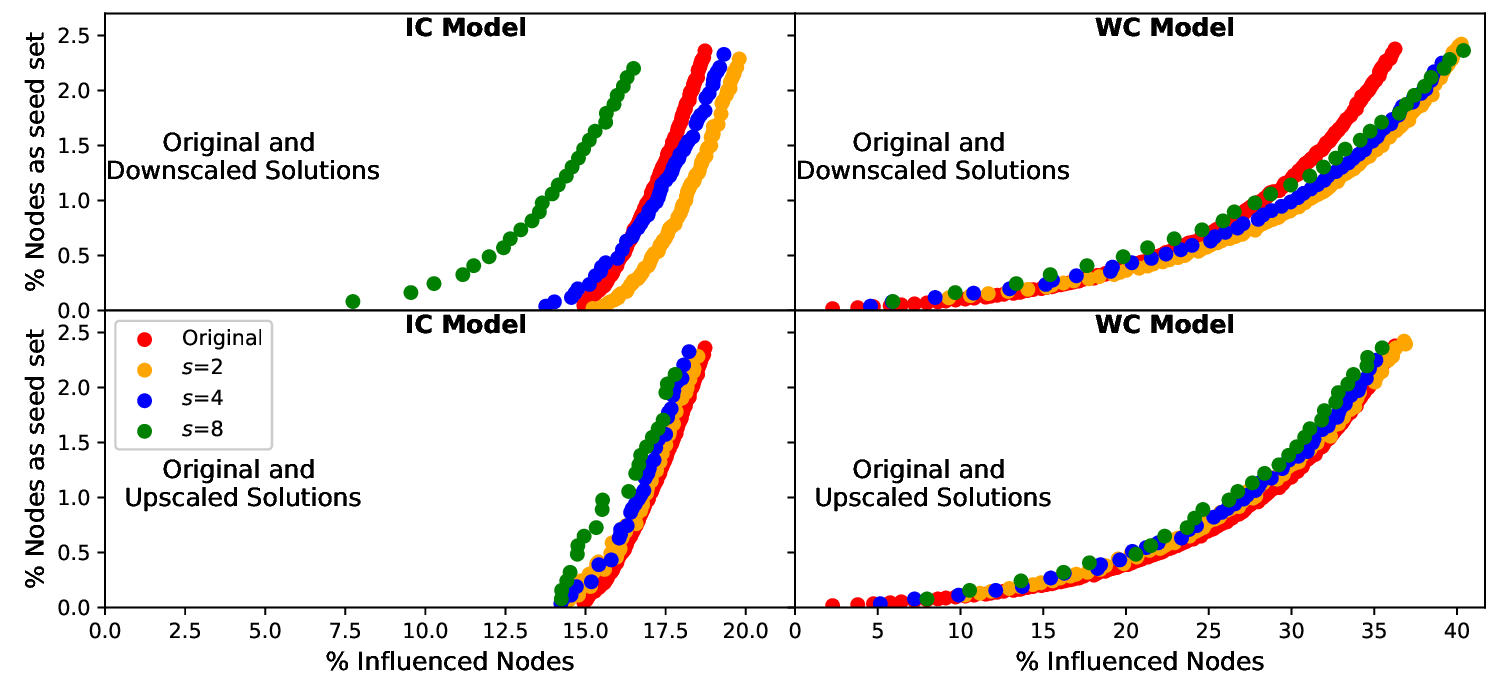}
	\caption{Fb. Pag. Pareto fronts: (top) for the original and downscaled networks, and (bottom) for the original network and the upscaled solutions.}
\label{fig:PF2}
\vspace{-0.5cm}
\end{figure}

Of note, the PGP case shown in Fig.~\ref{fig:PF1} is actually one of the cases where our method performs worst (see below). One of the cases where it performs best is Fb. Pag., whose PFs are shown in Fig.~\ref{fig:PF2} (also in this case, for one run of the MOEA on the original network and on the downscaled networks, using PageRank for upscaling). On this network, the MOEA on the downscaled networks often produces a better PF than on the original (top in Fig.~\ref{fig:PF2}), and the final, upscaled PFs are comparable to the original one (bottom in Fig.~\ref{fig:PF2}). Thus, the method largely preserved the quality of the solutions.

In general and quantitatively, we observe great variation among the networks under test in terms of HR, i.e., on the ratio between (1) the HV subtended by the PF found by the MOEA on the original network and (2) the HV subtended by the PF obtained from the upscaled solutions.

\begin{figure}[ht!]
\vspace{-0.5cm}
\centering
	\includegraphics[width=\textwidth,clip,trim=4mm 4mm 14mm 4mm]{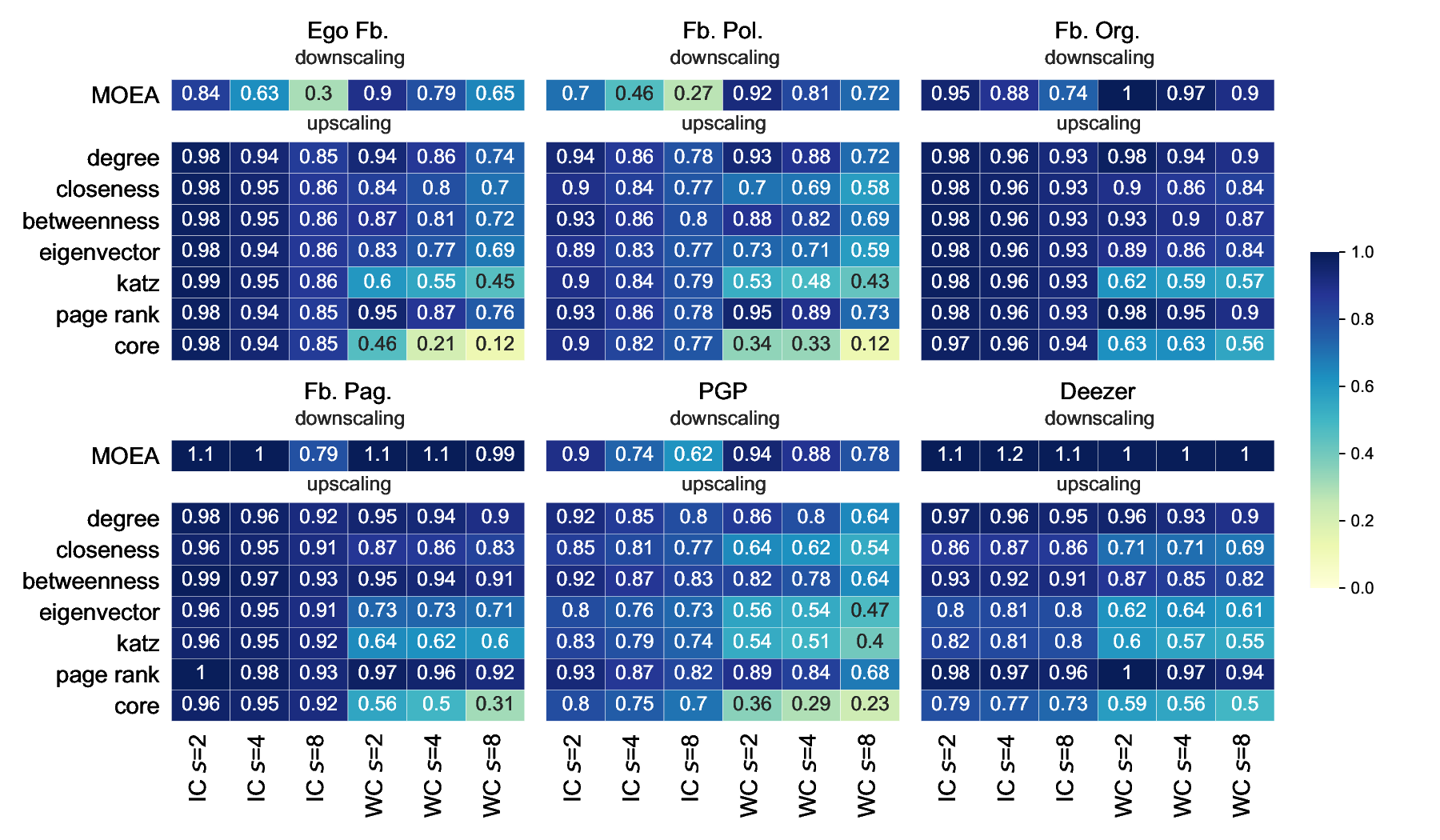}
	\caption{Hyperarea (averaged across 10 runs) for each network, propagation model, scaling factor, and centrality metric.}
\label{fig:HR}
\vspace{-0.4cm}
\end{figure}

To provide a more robust estimate of the HR values, for each network and propagation model we executed the MOEA in 10 independent runs on the original network, and in 10 runs for each value of scaling factor\footnote{We also compared directly the HV values. We applied the Wilcoxon Rank-Sum test (with $\alpha=0.05$) to analyse whether the HV values calculated on the downscaled solutions and the upscaled ones (with upscaling based on PageRank) were significantly different from the HV values obtained on the original network, all on 10 runs. All the pairwise comparisons resulted statistically significant, excluding the one related to the downscaled solutions found on Fb. Pag. with $s=2$ and WC model.}. We show the HR values (averaged across 10 runs) comparatively in Fig.~\ref{fig:HR}. Each cell contains the HR for a particular network, propagation model, and scaling factor. The ``MOEA'' rows contain intermediate HR values, which compare the PF on the {\bf downscaled} networks with the PF on the original network. For three out of six networks (Ego Fb., Fb. Pol., and PGP) the HR never reaches 1, meaning that the HV on the downscaled networks is \emph{lower} than the one on the original network. For the other three networks, HR reaches or surpasses 1, for at least some scaling factor and one or both propagation models, meaning that the downscaling step by itself preserved or even raised the optimality of the PF.

The HR values in the rows labelled with the various centrality names in Fig.~\ref{fig:HR} compare instead the PF of the {\bf upscaled} solutions with the PF on the original network, so serve as evaluation metrics for our method. These show how accurate each centrality is at the upscaling task, and thus help choose the most suitable centrality for any future work. We observe that it is generally easier for all centralities to obtain a good PF of upscaled solutions when using IC. However, three of the centralities consistently yield good upscaling results for both IC and WC: PageRank, betweenness, and degree centralities. PageRank is the best option overall (this is why we used it in Fig.~\ref{fig:PF1} and Fig.~\ref{fig:PF2}). Across networks with IC, the HR obtained by PageRank is in the interval $[0.93, 1]$ when $s=2$, in $[0.86, 0.97]$ when $s=4$, and in $[0.78, 0.96]$ when $s=8$. Fairly similar numbers are obtained with WC. 

\subsection{Runtime analysis}

\begin{figure}[ht!]
\vspace{-0.4cm}
\centering
	\includegraphics[width=\textwidth]{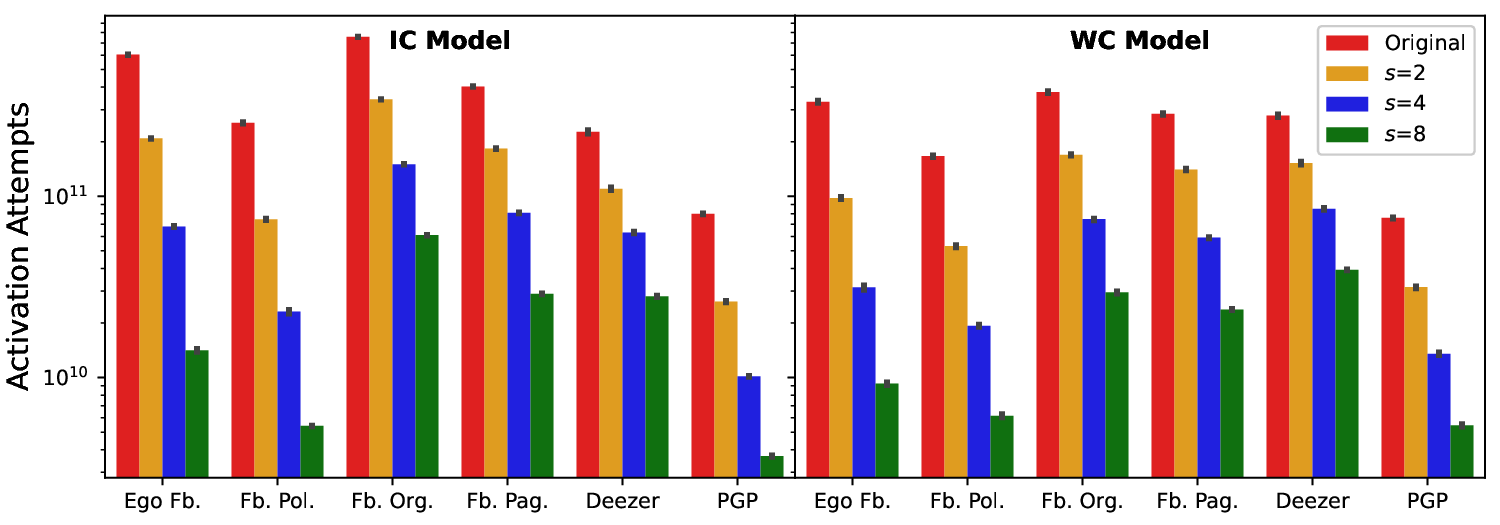}
	\caption{Runtime (no. activation attempts) in the comparative experiments with MOEA on the original network (average across $10$ runs, error bars indicate std. dev.).}
\label{fig:time_log}
\vspace{-0.9cm}
\end{figure}

The proposed method not only can preserve the quality of the solutions, but gains drastically in runtime due to our design based on downscaling the input data. We compute the runtime in terms of number of \emph{activation attempts} (line 7 in Algorithm~\ref{alg:cascade}). This is a proxy metric for the actual runtime, as it counts how many times that activation step is executed during all the simulations needed by a MOEA run (either on the original network, or on the downscaled one). We show these measurements in Fig.~\ref{fig:time_log}, from which we can see (note the log scale on the y-axis) that the runtime needed by our method decreases by a factor from two to five when the scaling factor is doubled.

\subsection{Comparison with heuristic algorithm}

To further prove the applicability of the proposed method, we conducted a final set of experiments on two different networks with $\sim$50k nodes: soc-gemsec \cite{rozemberczki2019gemsec} and soc-brightkite \cite{cho2011friendship}.
The goal of these experiments was to test our method with higher scaling factors on larger networks.

Due to the large runtime required, we executed only one run of the proposed method with the WC model for $s=16$ and $s=32$, and compared the results with those obtained by the deterministic greedy CELF algorithm \cite{Leskovec2007CosteffectiveOD} computed on the unscaled network.

\begin{figure}[ht!]
\vspace{-0.2cm}
\centering
	\includegraphics[width=\textwidth]{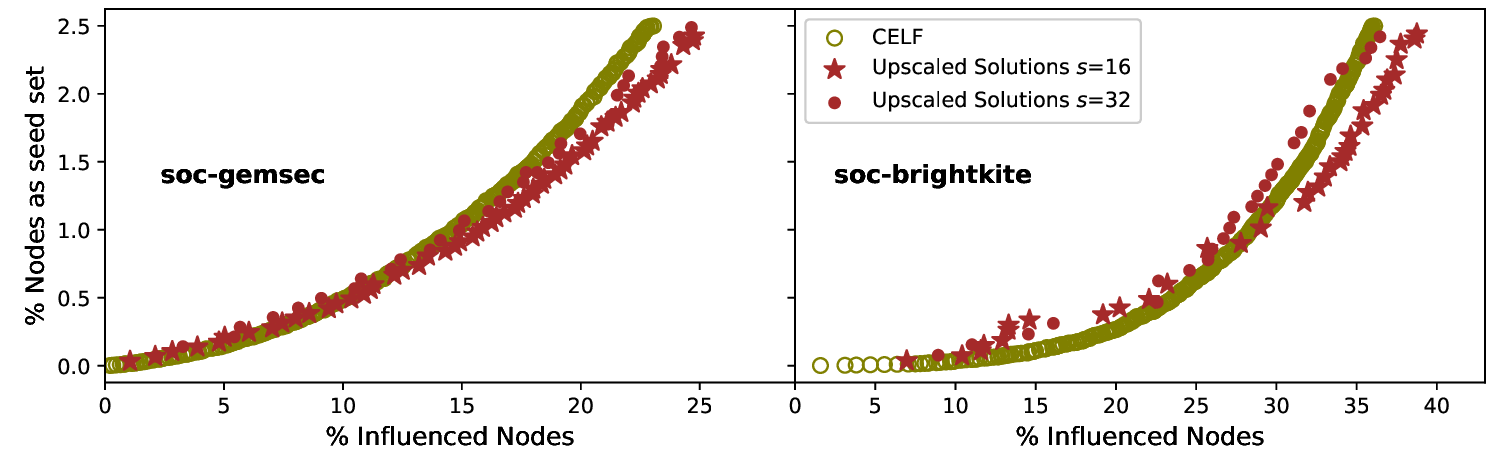}
	\caption{PFs obtained with CELF and with our method for $s=16$ and $s=32$.}
\label{fig:CELF}
\end{figure}

\begin{figure}[ht!]
\centering
	\includegraphics[width=\textwidth]{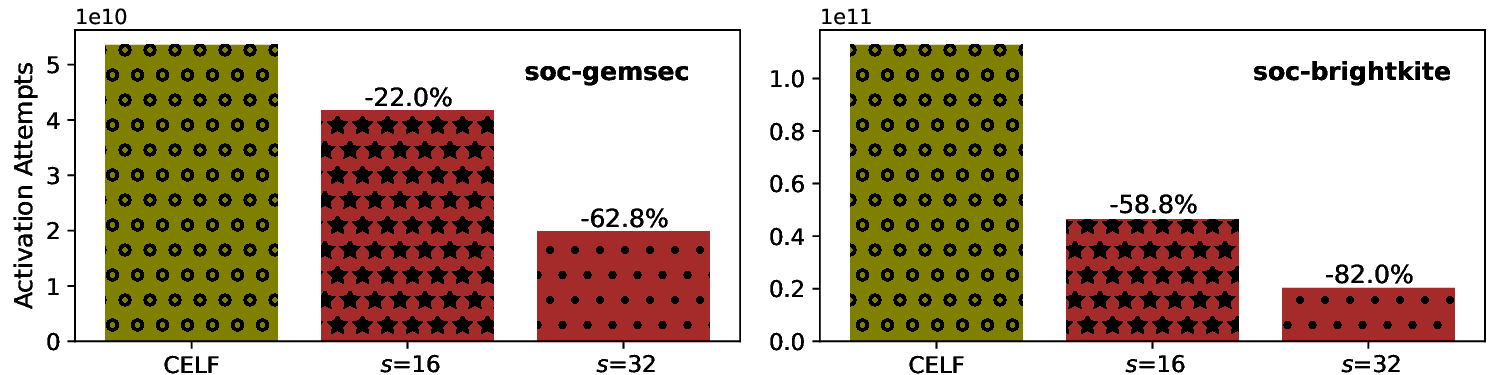}
	\caption{Runtime (no. activation attempts) measurements in the comparative experiment with CELF.}
\label{fig:Runtime_CELF}
\end{figure}

Despite the computational limitations, the comparison is still informative. We see in Fig.~\ref{fig:CELF} that our proposed method is able to achieve better results than CELF, even for these high scaling factors. Not only that: Fig.~\ref{fig:Runtime_CELF} shows that we obtain these results with a much lower number of activation attempts compared to CELF (with a reduction of up to $82\%$).

It is worth to mention that the results obtained with our method depend on the parameters of the MOEA, which remained the same as in the previous experiments, see Table~\ref{tab:params}.
However, we have noticed that the HV reaches a plateau after $\sim$300 generations. This means that with a simple implementation of a convergence termination criteria, the gain in runtime would be even higher.

\section{Discussion and conclusions}
\label{sec:discussion}

In this paper, we proposed a novel approach to tackle the IM problem where the focus is on the input network instead of the algorithm itself. For this reason, while here we tested the approach on a MOEA, the same method could be applied, in principle, to any other IM algorithm, such as the ones described in Section \ref{sec:intro}.

Our method has been proven to work correctly, although with some differences in the results, with both IC and WC propagation models on all the tested networks, regardless their sizes or properties.

The results demonstrate the effectiveness of our method in terms of quality of the solutions obtained. Furthermore, our method is able to drastically decrease, even for large networks, the runtime of the MOEA. The latter has the additional advantage of providing a whole set of diverse solutions (i.e., seed sets), unlike heuristic methods that are usually designed to return just one seed set.

Our method has proven to work properly with different scaling values. Nevertheless, the results show a lower quality of the solutions as the scaling factor increases. This result is in line with expectations, as it shows a clear trade-off between solution quality and runtime gain.

To the best of our knowledge, this is the first work where a downscaling and upscaling process is proposed on networks to solve the IM problem.
Regarding the centrality measure used in the upscaling process, we can state the best one to be PageRank, given the high quality of the upscaling results obtained and the low complexity required to compute it.

\clearpage

\bibliographystyle{splncs04}
\bibliography{bib-file}

\begin{thebibliography}{10}
\providecommand{\url}[1]{\texttt{#1}}
\providecommand{\urlprefix}{URL }
\providecommand{\doi}[1]{https://doi.org/#1}

\bibitem{boguna2004models}
Bogu{\~n}{\'a}, M., Pastor-Satorras, R., D{\'\i}az-Guilera, A., Arenas, A.:
  {Models of social networks based on social distance attachment}. Physical
  Review. E  \textbf{70}(5),  056122 (2004)

\bibitem{broido2019scale}
Broido, A.D., Clauset, A.: {Scale-free networks are rare}. Nature
  communications  \textbf{10}(1),  1--10 (2019)

\bibitem{bucur2020top}
Bucur, D.: {Top influencers can be identified universally by combining
  classical centralities}. Scientific Reports  \textbf{10}(1),  1--14 (2020)

\bibitem{Bucur2016InfluenceMI}
Bucur, D., Iacca, G.: {Influence Maximization in Social Networks with Genetic
  Algorithms}. In: {EvoApplications} (2016)

\bibitem{bucur2018improving}
Bucur, D., Iacca, G., Marcelli, A., Squillero, G., Tonda, A.: {Improving
  multi-objective evolutionary influence maximization in social networks}. In:
  {EvoApplications}. pp. 117--124 (2018)

\bibitem{Bucur2017MultiobjectiveEA}
Bucur, D., Iacca, G., Marcelli, A., Squillero, G., Tonda, A.P.:
  {Multi-objective Evolutionary Algorithms for Influence Maximization in Social
  Networks}. In: {EvoApplications} (2017)

\bibitem{cho2011friendship}
Cho, E., Myers, S.A., Leskovec, J.: Friendship and mobility: user movement in
  location-based social networks. In: Proceedings of the 17th ACM SIGKDD
  international conference on Knowledge discovery and data mining. pp.
  1082--1090. ACM (2011)

\bibitem{Deb2002AFA}
Deb, K., Agrawal, S., Pratap, A., Meyarivan, T.: {A fast and elitist
  multiobjective genetic algorithm: NSGA-II}. IEEE Transactions on Evolutionary
  Computing  \textbf{6},  182--197 (2002)

\bibitem{Fire2013OrganizationMU}
Fire, M., Puzis, R., Elovici, Y.: {Organization Mining Using Online Social
  Networks}. Networks and Spatial Economics  \textbf{16},  545--578 (2013)

\bibitem{Goldenberg2001TalkOT}
Goldenberg, J., Libai, B., Muller, E.: {Talk of the Network: A Complex Systems
  Look at the Underlying Process of Word-of-Mouth}. Marketing Letters
  \textbf{12},  211--223 (2001)

\bibitem{Gong2016AnEM}
Gong, M., Song, C., Duan, C., Ma, L., Shen, B.: {An Efficient Memetic Algorithm
  for Influence Maximization in Social Networks}. IEEE Computational
  Intelligence Magazine  \textbf{11},  22--33 (2016)

\bibitem{Gong2016InfluenceMI}
Gong, M., Yan, J., Shen, B., Ma, L., Cai, Q.: {Influence maximization in social
  networks based on discrete particle swarm optimization}. Information Sciences
   \textbf{367-368},  600--614 (2016)

\bibitem{Gney2021LargescaleIM}
G{\"u}ney, E., Leitner, M., Ruthmair, M., Sinnl, M.: Large-scale influence
  maximization via maximal covering location. Eur. J. Oper. Res.  \textbf{289},
   144--164 (2021)

\bibitem{Holland1983StochasticBF}
Holland, P., Laskey, K.B., Leinhardt, S.: {Stochastic blockmodels: First
  steps}. Social Networks  \textbf{5},  109--137 (1983)

\bibitem{iacca2021evolutionary}
Iacca, G., Konotopska, K., Bucur, D., Tonda, A.: {An evolutionary framework for
  maximizing influence propagation in social networks}. Software Impacts
  \textbf{9},  100107 (2021)

\bibitem{Jiang2011SimulatedAB}
Jiang, Q., Song, G., Cong, G., Wang, Y., Si, W., Xie, K.: {Simulated Annealing
  Based Influence Maximization in Social Networks}. In: {AAAI} (2011)

\bibitem{Karrer2011StochasticBA}
Karrer, B., Newman, M.E.J.: {Stochastic blockmodels and community structure in
  networks}. Physical Review. E  \textbf{83 1 Pt 2},  016107 (2011)

\bibitem{kempe2003maximizing}
Kempe, D., Kleinberg, J., Tardos, {\'E}.: {Maximizing the spread of influence
  through a social network}. In: {KDD}. pp. 137--146 (2003)

\bibitem{Konotopska2021GraphawareEA}
Konotopska, K., Iacca, G.: {Graph-aware evolutionary algorithms for influence
  maximization}. In: GECCO Companion (2021)

\bibitem{Leskovec2007CosteffectiveOD}
Leskovec, J., Krause, A., Guestrin, C., Faloutsos, C., Vanbriesen, J.M.,
  Glance, N.S.: {Cost-effective outbreak detection in networks}. In: {KDD}
  (2007)

\bibitem{Li2019TipTopE}
Li, X., Smith, J.D., Dinh, T.N., Thai, M.T.: Tiptop: (almost) exact solutions
  for influence maximization in billion-scale networks. IEEE/ACM Transactions
  on Networking  \textbf{27},  649--661 (2019)

\bibitem{Li2018InfluenceMO}
Li, Y., Fan, J., Wang, Y., Tan, K.L.: Influence maximization on social graphs:
  A survey. IEEE Transactions on Knowledge and Data Engineering  \textbf{30},
  1852--1872 (2018)

\bibitem{Liu2016GraphSM}
Liu, Y., Safavi, T., Dighe, A., Koutra, D.: Graph summarization methods and
  applications: A survey. arXiv: Information Retrieval  (2016)

\bibitem{lotf2022improved}
Lotf, J.J., Azgomi, M.A., Dishabi, M.R.E.: {An improved influence maximization
  method for social networks based on genetic algorithm}. Physica A:
  Statistical Mechanics and its Applications  \textbf{586},  126480 (2022)

\bibitem{ma2022influence}
Ma, L., Shao, Z., Li, X., Lin, Q., Li, J., Leung, V.C., Nandi, A.K.: {Influence
  Maximization in Complex Networks by Using Evolutionary Deep Reinforcement
  Learning}. IEEE Transactions on Emerging Topics in Computational Intelligence
   (2022)

\bibitem{McAuley2012LearningTD}
McAuley, J., Leskovec, J.: {Learning to Discover Social Circles in Ego
  Networks}. In: {NIPS} (2012)

\bibitem{newman2018networks}
Newman, M.: {Networks}. Oxford University Press (2018)

\bibitem{Peixoto2017NonparametricBI}
Peixoto, T.P.: {Nonparametric Bayesian inference of the microcanonical
  stochastic block model}. Physical Review. E  \textbf{95 1-1},  012317 (2017)

\bibitem{Richardson2003TrustMF}
Richardson, M., Agrawal, R., Domingos, P.M.: {Trust Management for the Semantic
  Web}. In: {SEMWEB} (2003)

\bibitem{rozemberczki2019gemsec}
Rozemberczki, B., Davies, R., Sarkar, R., Sutton, C.: {GEMSEC: Graph Embedding
  with Self Clustering}. In: {ASONAM}. pp. 65--72 (2019)

\bibitem{rozemberczki2020characteristic}
Rozemberczki, B., Sarkar, R.: {Characteristic functions on graphs: Birds of a
  feather, from statistical descriptors to parametric models}. In: {CIKM}. pp.
  1325--1334 (2020)

\bibitem{Traag2019FromLT}
Traag, V.A., Waltman, L., van Eck, N.J.: {From Louvain to Leiden: guaranteeing
  well-connected communities}. Scientific Reports  \textbf{9} (2019)

\bibitem{wang2022multi}
Wang, C., Zhao, J., Li, L., Jiao, L., Liu, J., Wu, K.: A multi-transformation
  evolutionary framework for influence maximization in social networks. arXiv
  preprint arXiv:2204.03297  (2022)

\bibitem{Wang2012ScalableIM}
Wang, C., Chen, W., Wang, Y.: {Scalable influence maximization for independent
  cascade model in large-scale social networks}. Data Mining and Knowledge
  Discovery  \textbf{25},  545--576 (2012)

\bibitem{Wu2018ATS}
Wu, H.H., K{\"u}ç{\"u}kyavuz, S.: A two-stage stochastic programming approach
  for influence maximization in social networks. Computational Optimization and
  Applications  \textbf{69},  563--595 (2018)

\bibitem{zitzler1998multiobjective}
Zitzler, E., Thiele, L.: Multiobjective optimization using evolutionary
  algorithms—a comparative case study. In: International conference on
  parallel problem solving from nature. pp. 292--301. Springer (1998)

\end{thebibliography}


\end{document}